\documentclass{aa}
\usepackage{graphicx}
\usepackage{natbib}
\usepackage{txfonts}
\usepackage{epstopdf}
\usepackage{hyperref}
\bibpunct{(}{)}{;}{a}{}{,}    % to follow the A&A style

\def\apjs{ApJS}

\def\grl{Geophys. Res. Lett.}
\def\aap{Astron. Astrophys.}
\def\apj{Astrophys. J.}
\def\solphys{Sol. Phys.}
\def\nat{Nature}
\def\mnras{Mon. Not. R. Astron. Soc.}

\begin{document}
\title{Correlation of Spectral Solar Irradiance with solar activity as measured by SPM/VIRGO}
\author{C. Wehrli \and W. Schmutz \and A. I. Shapiro }
\offprints{C. Wehrli}

\institute{Physikalisch-Meteorologisches Observatorium Davos, World Radiation Center, 7260 Davos Dorf, Switzerland\\
\email{christoph.wehrli@pmodwrc.ch}}
\date{Received ; accepted }

\abstract
% Context (optional)
{The variability of Solar Spectral Irradiance over the rotational period and its trend over the solar activity cycle are important for understanding the Sun--Earth connection as well as for observational constraints for solar models. Recently the SIM experiment on SORCE has published an unexpected negative correlation with Total Solar Irradiance of the visible spectral range. It is compensated by a strong  and positive variability of the near UV range.}
% aims heading (mandatory)
{We aim to verify whether the anti-correlated SIM/SORCE-trend in the visible can be confirmed by independent observations of the VIRGO experiment on SOHO. The challenge of all space experiments measuring solar irradiance are sensitivity changes of their sensors due to exposure to intense UV radiation, which are difficult to assess in orbit.}
% methods heading (mandatory)
{We exclude the first 6 years prior to 2002 where one or more fast processes contributed to instrumental changes and analyze a 10-year timeseries of VIRGO sun photometer data between 2002 and 2012. The variability of Spectral Solar Irradiance is correlated with the variability of the Total Solar Irradiance, which is taken as a proxy for solar activity.}
% results heading (mandatory)
{Observational evidence indicates that after six years only one single long-term process governs the degradation
of the backup sun photometer in VIRGO which is operated once in a month. This degradation can be well approximated by a linear function over ten years. The analysis of the residuals from the linear trend yield robust positive correlations of spectral irradiance at 862, 500 and 402\,nm with total irradiance. In the analysis of annual averages of these data the positive correlations change into  weak negative correlations, but of little statistical significance, for the 862\,nm and 402\,nm data. At 500\,nm the annual spectral data are still positively correlated with Total Solar Irradiance. The persisting positive correlation at 500 nm is in contradiction to the SIM/SORCE results.}
% conclusions heading (optional)
{}


\titlerunning{SPM/VIRGO correlation with solar activity}

\maketitle
%
%_____________________________________________________________________
\section{Introduction}\label{sect:intro}

The variability of the solar irradiance  was suspected long ago \citep{smyth1855}, but its connection with solar activity was not clear till the beginning of the satellite measurements of the Total Solar Irradiance (TSI) in 1978. The measurements showed that TSI is positively correlated with solar activity, so generally the more sunspots are visible in the Sun, the larger is the TSI value. The mean level of the TSI during the solar maximum is approximately 0.1 \% higher than during the solar minimum.  This can be successfully explained by the current models \citep{leanetal2005, krivovasolanki2008, 2011Sci...334..916S}, which attribute the solar variability on the 11-year time scale to the imbalance between the contributions from dark sunspots and bright active features (i.e. plage and bright network).

The understanding of the Spectral Solar Irradiance (SSI) variability is significantly hampered by the lack of the long-term measurements. The contrast between the bright features and quite Sun decreases with the wavelengths as the Planck function becomes less sensitive to the temperature changes \citep{shapiroetal2010}. Therefore, most of the models \citep[e.g.][]{leanetal2005,krivovasolanki2008} predict inverse solar cycle changes (i.e. anticorrelation between the solar activity and irradiance) in the IR, with the inversion point located in between 1200 and 1600 nm. The UV and visible irradiance are predicted to positively correlate with the solar cycle (see review by \cite{ermolliealt2012} for detailed intercomparison of the available models).

Strikingly,  the recent SIM/SORCE measurements showed completely different picture of the solar variability in the UV and visible \citep{harderetal2009}. The variability in UV measured by SIM/SORCE is several times higher than all recent estimates. While 11-year variability of the UV and IR irradiance are in phase with solar cycle, the variability of the visible irradiance (400 - 730 nm) is in antiphase. It was suggested that such behavior may be caused by a negative correlation of the continuum intensity with magnetic field observed at some wavelengths \citep{topkaetal1997} or by a change of the temperature gradient in the photosphere (see \cite{harderetal2009,fontenlaetal2011} for more detailed discussion).

Presently there is an evidence both in favour and against the anticorrelation of the visible irradiance and activity reported by \cite{harderetal2009}.   \citet{premingeretal2011} showed that the analysis of the full-disk photometric images from the San Fernando Observatory indicates the opposite trends in the TSI and visible, which is in agreement with SIM measurements. At the same time \cite{leandeland2012} suggested that SIM measurements are affected by the uncorrected sensitivity drifts.
The clarification of this question is of crucial importance for the climate modeling \citep{haighetal2010, shapiroetal2013}.

In this paper we study the measurements of the SPM-B filterradiometer in VIRGO on SOHO searching for a qualitative statement about long-term correlation of SSI with solar activity between 2002 and 2012. In Sect.~2 we describe some technical background of the instrument, its operation and in-flight performance that we deem necessary for understanding of the data set. Sect.~3 addresses our approach to separate solar cycle variability from instrumental degradation and provides the results of straight forward correlation with and regression versus TSI observations. A summary and conclusion are given in Sect.~4.

\begin{figure}
\centering
%\includegraphics{Fig1A.eps}
%\includegraphics{Fig1B.eps}
%\caption{Time series of SPM-A (left panel) and SPM-B  (right panel) spectral irradiance at 862nm (red), 500nm %(green), and 402nm (blue)}
\includegraphics{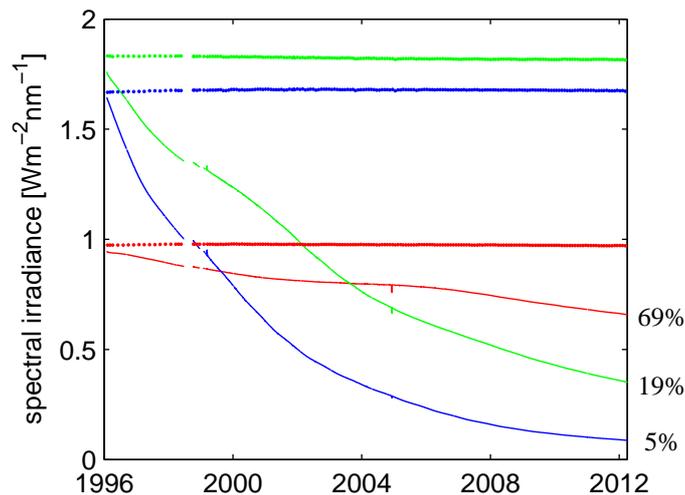}
\caption{Time series of SPM-A (line) and SPM-B  (dots, horizontal) spectral irradiance at 862nm (red), 500nm (green), and 402nm (blue)}
\label{fig:degr}
\end{figure}

\section{Instrument description}\label{sect:VIRGO}
The VIRGO SPM is a spectral radiometer with three channels centered at 402, 500, and 862 nm using interference filters of 5 nm FWHM bandwidth and silicon detectors. The filters are protected from energetic particle flux and UV radiation below 380 nm by a 3 mm thick front window of radiation resistant BK7-G18 glass.
The optical compartment containing the filters and detectors is physically separated from the rest of the VIRGO experiment, and was permanently flushed with 5.5 grade nitrogen until launch. Further details about the VIRGO experiment are given in \cite{Frohlich1995}.

\subsection{Operations}
Both heaters of the SPM instruments were activated with covers still closed on 6 December 1995 for an outgassing period of 46 days. The filter and detector compartments are constantly heated to a few degrees above ambient satellite temperature in order to reduce condensation of gaseous contaminants on their optical surfaces.

The cover of SPM-A was opened on 17 January 1996 and this instrument has since then continuously observed the Sun for more than 5900 days with the exception of the so called ``SOHO vacation'' from June 25 to September 25, 1998. This covers the complete solar cycle 23 and the onset of cycle 24.

SPM-A is used primarily as a helioseismological instrument and therefore operated continuously with a 60 sec sampling rate. SPM-B is the radiometric instrument, intended to monitor the solar spectral flux and the degradation of SPM-A. During backup operations it is exposed for 20 minutes taking 6 samples of 60 seconds for each of its 3 channels sequentially. Until end of 1998 backup measurements were taken every 60 days, and since February 1999 the backup cadence was doubled to 30 days.

Level-1 are fully processed data with all {\it a priori} known corrections, e.g. conversion to physical units W\,m$^{-2}$nm$^{-1}$, temperature corrections, and normalization to Sun--Earth distance of 1 AU applied. Daily means are calculated for SPM-A, while SPM-B data represent averages over 6 minutes.

\subsection{Instrument degradation}
Since the start of operational measurements in January 1996 the SPM-A has suffered a steady loss of sensitivity. After more than 15 years of continuous exposure the red channel signal is reduced to 70\%, the green to 20\%, and the blue to 5\% of their initial values. Several processes may contribute to this degradation: contamination of the first optical surface, deterioration of filter glass and interference coatings, loss of efficiency in the silicon photo detectors. SPM-B has maintained more than 99\% of its sensitivity (see Figure ~\ref{fig:degr}) due to its sparse operation cycle with an accumulated exposure time of less than 3 days versus more than 5900 days for SPM-A. Experience gained from similar instruments retrieved from the EURECA mission (\cite{Wehrli1995}) led to a much improved performance of the VIRGO spectral radiometer. Further speculations about the contribution of different processes leading to the sensitivity loss and their evolution in time are beyond the scope of this article.

\begin{figure}
\resizebox{\hsize}{!}{\includegraphics{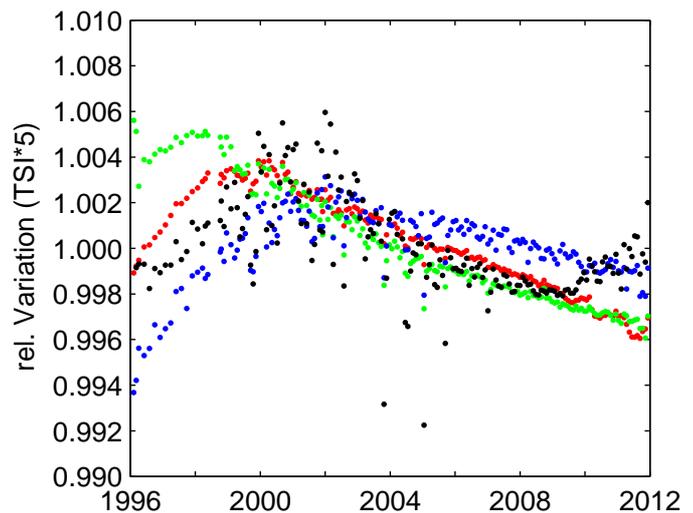}}
\caption{Variations relative to their mean of VIRGO spectral (blue, green, and red), and total solar irradiance (black) during SPM-B backup measurements. The TSI relative variations are expanded by a factor 5.}
\label{fig:B_original}
\end{figure}

\begin{figure}
\resizebox{\hsize}{!}{\includegraphics{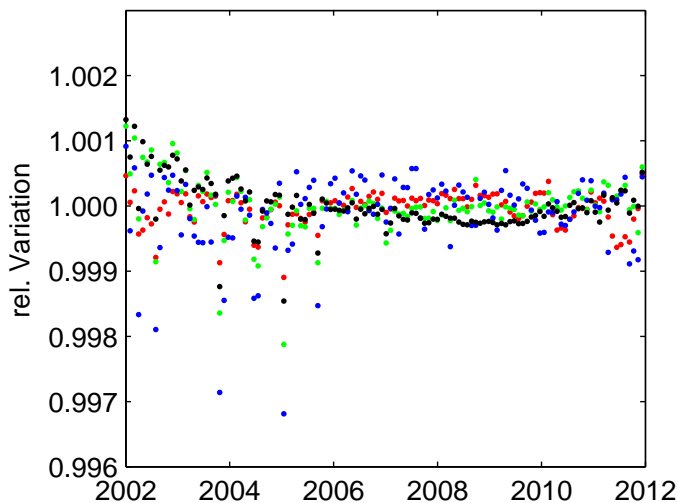}}
\caption{VIRGO spectral solar irradiance variations (red, green, blue) normalized to their linear trends and total solar irradiance (black) normalized to its mean value}
\label{fig:B_corr}
\end{figure}

\section{Variability of VIRGO spectral irradiance}\label{sect:Variability}

The VIRGO SPM time series represent the product of solar cycle variability and instrumental sensitivity changes that is difficult to factorize. The sensitivity rate of change is intricately related to solar activity as solar UV and energetic particles fluxes act on optical materials and semiconductor components to a largely unknown extent.
Depending on the approach taken to fit an empirical model for the degradation through the observations, the solar cycle variability may be eliminated completely or be grossly overestimated.

The apparent stability of SPM-B signal suggests that an analysis of spectral solar irradiance variability over the solar cycle is possible. Figure \ref{fig:B_original} shows the relative variations of fluxes measured by SPM-B together with the variation of the TSI channel measured at corresponding dates, but magnified by a factor of five.

The 2\,$\sigma$ variability in SSI observed by SPM-B between 1996 and 2009 is  around 4\,-\,5$\cdot10^{-3}$ or about 5 times larger than the corresponding TSI variation. From the general shape of the SSI variations common to all 3 channels, one would suspect an instrumental effect comprising at least two processes with different time constants, which can be modeled as the sum of 2 exponentials. Time constants of a bi-exponential fitted to the red and blue channel amount to 2\,--\,3 years for the fast and more than 1000 years for the slow component.  {The bi-exponential for doesn't fit the green channel well, likely due to a stronger solar component.}

Arguing that the ``fast'' processes of the instrumental effect has become insignificant after 2 time constants, i.e. after 2002, and that the ``slow'' component may be approximated by a linear function over a small fraction (1\% of 1000y) of its characteristic time, we propose to model the instrumental degradation since 2002 by a simple linear trend.  {Including data prior to 2002 would require rather accurate modeling of the 'fast' instrumental effects, which is beyond the scope of this paper.}

\begin{figure*}
\centering
\resizebox{0.3\hsize}{!}{\includegraphics{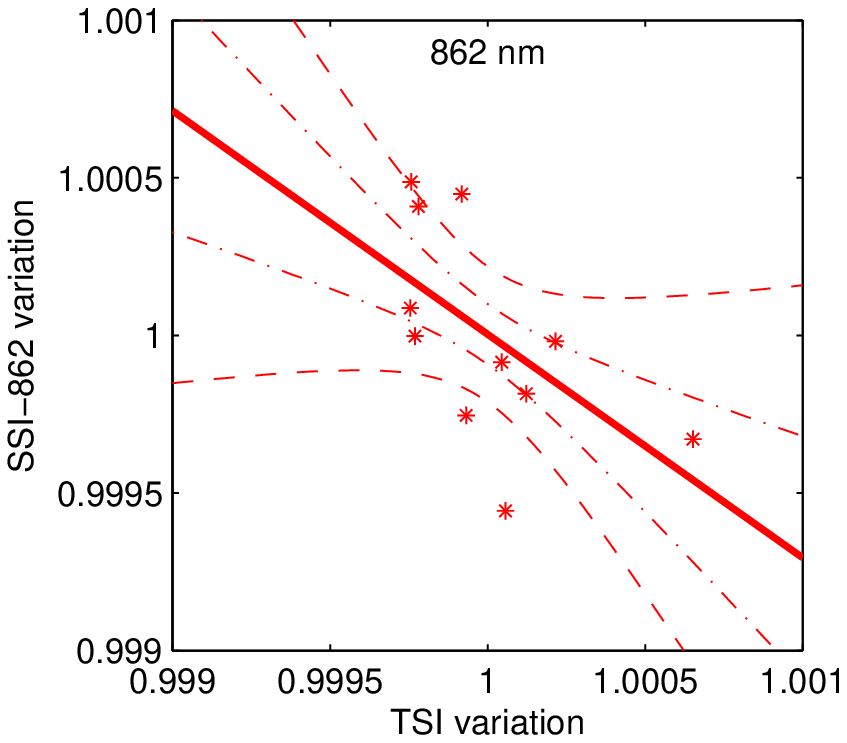}}
\resizebox{0.3\hsize}{!}{\includegraphics{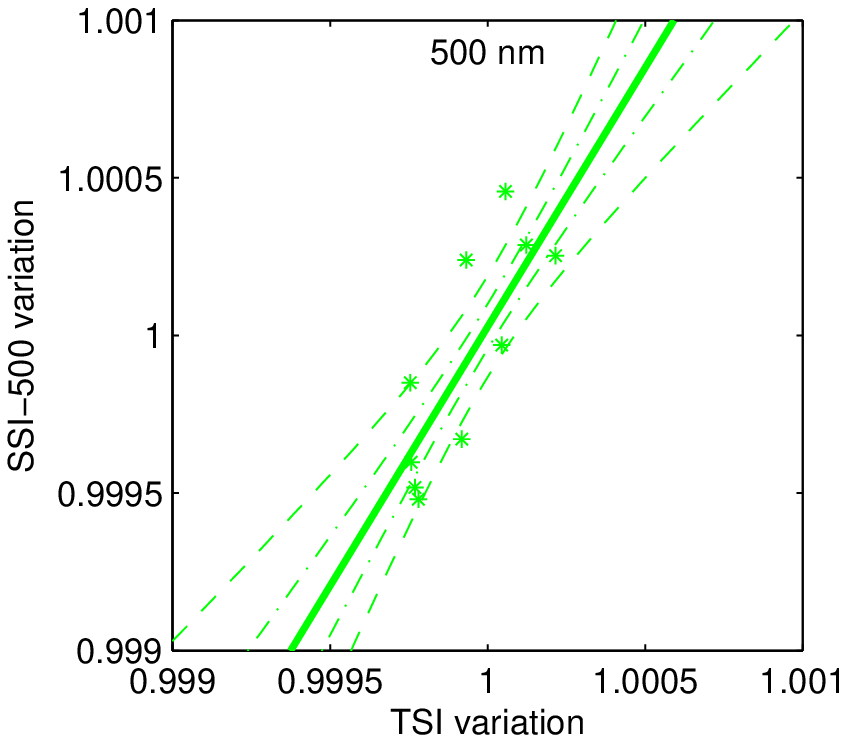}}
\resizebox{0.3\hsize}{!}{\includegraphics{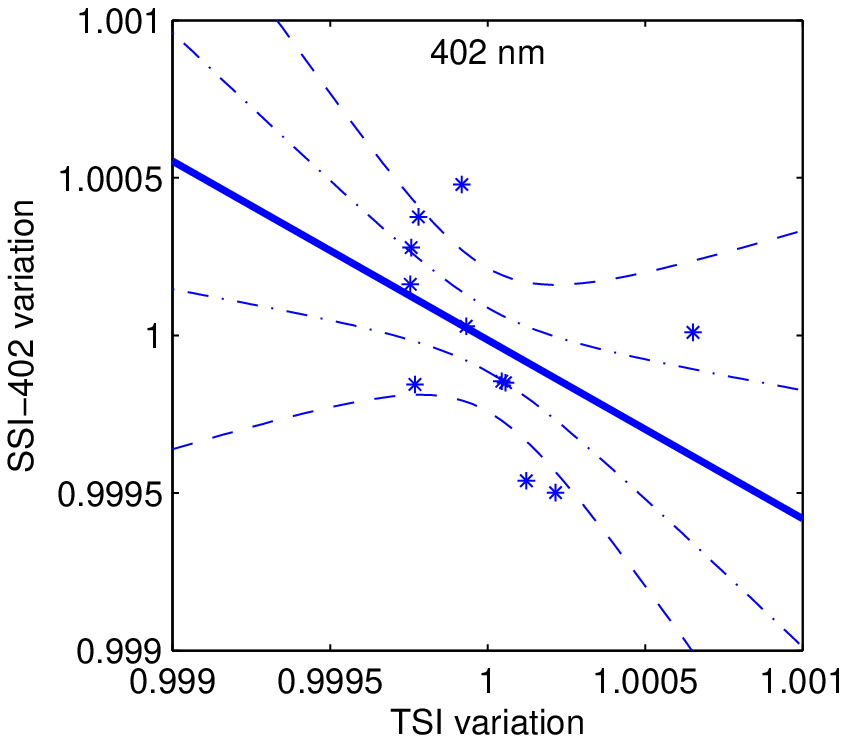}}
\caption{VIRGO SSI (RGB) versus TSI variations based on annual means from 2002 to 2012. Solid lines indicate the robust linear regressions, the dash dot and dashed lines are respectively 1\,$\sigma$ and 2\,$\sigma$ uncertainties.}
\label{fig:regr_ann}
\end{figure*}

The signature of SSI variability during the decline of solar cycle 23 and rise of cycle 24 can be represented by the residuals of a linear function fitted to the SPM-B data and be related to the variability of TSI  {according to PMOD composite \citep{PMODcomp} }.   {By our modest approach we eliminate most of the instrumental degradation and obtain an imperfect estimate of the solar variability.}
A statistical analysis of the spectral variability can then provide observational verification as to whether the anti-correlation of spectral versus total irradiance in the mid-visible wavelength range reported by \cite{harderetal2009} and \cite{premingeretal2011} can be confirmed by VIRGO measurements.

Figure~\ref{fig:B_corr} shows the variation of SSI since 2002 normalized to a robust linear least square fit \citep{Robust_reg} over the period shown. The 2\,$\sigma$ variability is now in the order of 5$\cdot 10^{-4}$ and comparable to the TSI variation show in black. Solar activity nicely shows up on rotational timescale, e.g. on 15 January 2005 with relative depressions of 0.9990 (Red), 0.9985 (TSI), 0.9979 (Green), and 0.9968 (Blue).
The distinct signature of declining cycle 23 and rising cycle 24 seen in TSI variations is discernible in the detrended variation of the green (500\,nm) channel, but is not obvious in the red (862\,nm) and blue (402\,nm) channels.

Table~\ref{table:daily} (columns 2--4) lists the correlation coefficients of 124 simultaneous observations of solar spectral and total irradiance variations between January 2002 and March 2012.
Numbers in brackets indicate the 95\% confidence intervals of the correlation. The statistically robust correlation for the visible channel is further illustrated by Figure ~\ref{fig:regr_daily} (shown in Online Material).
From these correlations can be concluded that: a) All spectral channels show a positive correlation with TSI, b) Correlations among the spectral channels are also positive, c)  The correlation of the 500\,nm channel with TSI is quite strong and robust as already seen in Figure 3, and d) The confidence intervals of the weaker correlation for the channels at 402\,nm and 862\,nm are still entirely positive.

 \begin{table*}
 \caption{Correlation coefficients between relative variations in TSI and SSI channels for monthly snapshot values (columns 2--4) and for annual averages (columns 5--7) are listed with 95\% uncertainties in brackets. The number of monthly and annual values are 124 and 11. }\label{table:daily}  \begin{center}
 \begin{tabular}{  c  c  c  c | c   c  c}
\hline
 & & Monthly snapshots & & & Annual averages & \\
 &        SSI-862    & 	SSI-500 	   &  SSI-402    &   SSI-862    & 	 SSI-500 	   &  SSI-402  \\
\hline
 TSI         	    	  &   0.25[0.08,0.41]	          & 0.87[0.83,0.91]		   & 0.48[0.33,0.60]     &   -0.58[-0.88,0.03]	          &	0.91[0.68,0.98]	   &    -0.43[-0.82,0.23]   \\
 SSI-862 		                &   1.0	 &  	0.45[0.30,0.58]    & 	  0.55[0.42,0.66]   &   1.0	 &  	 -0.79[-0.94,-0.36]   & 	  0.65[0.09,0.90]         \\
 SSI-500		              &    	 &	1.0   & 	  0.68[0.58,0.77]	   &	 &	 1.0   & 	  -0.47[-0.82,0.24]     \\
  SSI-402 		       &       &      &  1.0   &  &      &  1.0  \\
   \hline
\end{tabular}
\end{center}
\label{table:daily}
\end{table*}

\subsection{Correlation of annual means}\label{sect:annual}
Due to the sampling rate of 30 days for the SPM-B instrument, the positive correlation found in individual measurements represents a mixture of solar cycle and solar rotation activity.
In order to isolate the solar cycle trend from contamination by rotating solar activity, we analyzed annual means of SPM-B measurements from 2002 to 2012 (see Fig~\ref{fig:val_annual} in Online Material). In contrast to the all positive monthly correlations the annual correlations (see Table~\ref{table:daily}, columns 5--7) become negative in red and blue channels. Corresponding p-values,  {indicating} the probability that the given correlations might occur by  {chance}, are 0.06 for 862nm, \textless 0.01 for 500nm, and 0.19 for 402nm.
 {The sign of correlations remain within the uncertainties given in Table~\ref{table:daily} when the data set is restricted to shorter, e.g. 2002--2010, 2004--2012, or 2004--2010, periods.}

This finding is further corroborated by robust linear regressions of relative SSI versus TSI variations shown in Figure ~\ref{fig:regr_ann}. A statistically significant, positive slope of regression is thus found for the mid VIS channel. Interestingly enough, both the near IR and near UV channels show negative, though less significant, slopes given in Table~\ref{table:slopes} (given in Online Material).

\section{Discussion and Conclusions}
We have analyzed a time series of solar spectral irradiance measured by the VIRGO experiment on the SOHO satellite in terms of spectral irradiance variation versus simultaneous measurements of the total solar irradiance used as proxy for solar activity. Monthly backup measurements of the SPM-B instrument are affected by small, but not negligible, instrumental ageing effects masking the signature of solar activity. We note that there are several processes (e.g. shift of the peak wavelengths of the filters due to the diffusion of the filter's layers, developing of spectral leaks, radiation damage of the silicon detector, etc.) which could influence the ageing and, in principle, would make it a non monotonic function of time. Presently there is no reliable way to quantitatively characterize these processes or even to confirm their existence in the VIRGO/SPM data. Therefore the only way to take these potential effects into account is to introduce a sophisticated empirical model of the degradation with a large number of free parameters. By adjusting these free parameters one may control the share of instrumental effects and solar signature in the observed variability even to the extreme case of attributing the full variability to either of the two possibilities. We note that the same is true for all available SSI measurements and there are a number of examples when different analyses of the same SSI measurements led to strikingly different results (see e.g. review by Ermolli et al. 2013).

SPM-B daily averages were also analyzed by Fr{\"o}hlich\footnote{Poster at SORCE meeting 2011, available from  \url{ftp://ftp.pmodwrc.ch/pub/Claus/SORCE_2011/SPM_poster_cf.pdf}}
using a sophisticated model of instrument long-term behavior with 13 empirical parameters involving multiple time- and temperature dependencies, as well as UV irradiance and TSI used as surrogate for solar activity effects. He derives positive correlations at all three wavelengths, with cycle amplitude ratios of 0.5, 1.6 and 2.7 for the red, green and blue channel with respect to TSI. Such a strong  {positive} correlation would imply an even stronger negative correlation in the infrared tail of the spectrum in order to make the integrated SSI variability consistent with the observed TSI variability.

Here, we thus explicitly refrain from sophisticated speculations about the nature and strength of these instrumental effects and search for the least number of free parameters. Observational data indicate that the VIRGO ageing may be summarized by a 'fast' (2 - 3 years) and a 'slow' ($\approx$1000 y) processes, exponential in time, and that the 'fast' processes have decayed after 6 years, the 10 years period from January 2002 to March 2012 was empirically approximated by a linear function in time.

Based on monthly snapshots, positive correlations of SSI with TSI were found at all 3 wavelengths. In the analysis of annual averages, the correlations at 862 nm and at 402 nm become negative. A negative correlation for the IR channel is more likely real (R$^2$=0.22) than for the UV channel (R$^2$=0.01), but both are not significant at the 2-$\sigma$ level. However, the robust (R$^2$=0.79) positive correlation of monthly snapshots at 500 nm remains statistically significant at the 5-$\sigma$ level after smoothing out of the solar rotation period. This result based on VIRGO observations thus clearly contradicts previous reports \citep{harderetal2009} of anti-correlated spectral irradiance at the peak of the solar spectral energy distribution with the solar cycle.

\begin{acknowledgements}
These results would not be possible without the continuous efforts of the VIRGO and SOHO teams. SOHO is a cooperative ESA/NASA mission.
AS acknowledges the support from Swiss National Science Foundation under grant CRSI122-130642 (FUPSOL).
The work has also benefited from the travel support for the COST Action ES1005 TOSCA (http://www.tosca-cost.eu)
We thank Ga\"{e}l Cessateur for productive discussions and referee Jerald Harder for constructive criticism and helpful comments.
\end{acknowledgements}

\bibliographystyle{aa}
%\bibliography{shapiro}

\newpage

\Online

\begin{appendix}

\section{SSI variations at 500 nm versus TSI variations}

\begin{figure}
\resizebox{\hsize}{!}{\includegraphics{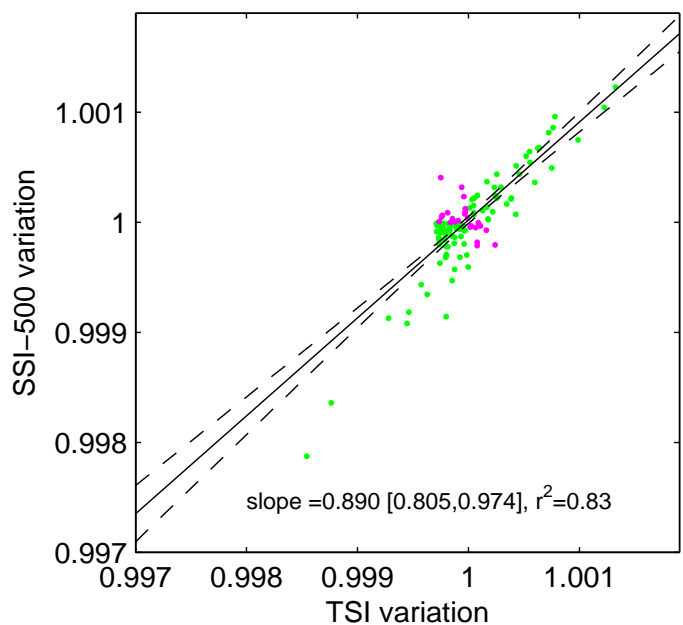}}
\caption{SSI variation at 500 nm versus TSI variation for the period from 2001 to 2012 with robust linear regression. Green dots (n=104) have positive correlation, magenta dots (m=21) are anti-correlated with TSI.}
\label{fig:regr_daily}
\end{figure}

A robust linear regression analysis of monthly snapshot SSI at 500\,nm versus TSI variations confirms the positive correlation reported here.
The lowest irradiance values occur during sun spot passages, the largest values during the active mid-term of cycle 23. Thus, data points shown in Figure~\ref{fig:regr_daily} lower left corner $(x, y <1)$ are representing measurements taken during passage of sunspots over the solar disc while those shown in upper right corner are representing observations during the declining phase of solar cycle 23 and rising phase of cycle 24. All data points are weighted by a bi-square function of their regression residues thereby the leverage of outliers, e.g. transit of large sun spots, on regression is minimized. We note that the regression slope is clearly dominated by the number of data points (green) with positive correlation, specifically by those in the upper right corner representing higher solar activity.

We note further that all points (magenta) which correspond to anti-correlation between the TSI and SSI-500 are located very close to (1,1) point on Figure~\ref{fig:regr_daily}. This means that all cases of anti-correlation occur when the amplitude of the cycle is small and comparable to the uncertainty of the degradation correction. Therefore, these points are the most sensitive to small deviations in the degradation correction and accordingly the least reliable. {When the data are restricted to the central cluster (within a square region of 1$\pm$0.0005), the slope is reduced to +0.561 [0.436, 0.686], but remains strictly positive.}

\begin{figure}
\resizebox{\hsize}{!}{\includegraphics{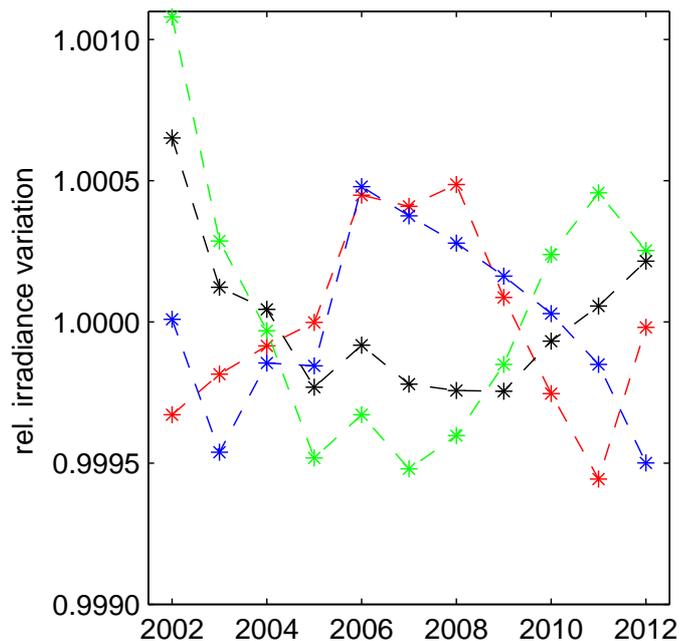}}\caption{The same as Fig.~\ref{fig:B_corr} but for annual mean values. }
\label{fig:val_annual}
\end{figure}

The trend of annual-mean data shown in Figure~\ref{fig:val_annual} obvious to eye: the green curve (500 nm) resembles the black (TSI) while the red (862 nm) blue (402 nm) line almost correspond to images of TSI mirrored at normalization level 1.0.
 {We note that the temporal position of the extrema of SSI variations may be affected by the choice of the selected period. However the sign of the correlation remains unaffected (see Sect~\ref{sect:annual})}.

\begin{table}
 \caption{Slopes of linear regression (shown in Figure ~\ref{fig:regr_ann})   between relative variations in VIRGO TSI and SSI annual means of monthly snapshots.}\label{table:slopes}
 \begin{center}
 \begin{tabular}{c  c  c  c  c}
\hline
  Channel   &        Slope       &   2\,$\sigma$-uncertainty    & 	 1\,$\sigma$-uncertainty  	   &  R$^2$    \\
\hline
 SSI-862 		  &     -0.71         &   [-1.55,\,0.13]	 &  [-1.08,\,-0.34] & 	  0.22       \\
 SSI-500		 &       1.65      &    [1.02, ,2.27]	 &	 [1.37,\,1.92]   & 	 0.79	      \\
  SSI-402 		&     -0.57  &  [-1.45,\,0.32]     &    [-0.96 ,\,-0.17]  &  0.01 \\
   \hline
\end{tabular}
\end{center}
\end{table}

\end{appendix}

\end{document}